\documentclass[12pt]{article}

\usepackage{sbc-template}

\usepackage{float}
\usepackage{graphicx,url}

\usepackage[brazil]{babel}   
\usepackage[T1]{fontenc}
\usepackage[latin1]{inputenc}  

\newcounter{algorithm}
\newenvironment{algorithm}[1]%
{\def\algcap{#1}%
\newdimen\quadamount\quadamount=1.5em%
\def\C##1##2{\\[.75ex]\hskip##1\quadamount{\bf \# ##2}}%
\def\CC##1{\hskip2\quadamount{\bf \# ##1}}%
\def\CCC##1{\hfill{\bf \# ##1}}%
\def\TAB##1{\\[.75ex]\hskip##1\quadamount\relax}%
\def\I{\TAB1}%
\def\II{\TAB2}%
\def\III{\TAB3}%
\begin{figure}[H]%
\hrulefill\par\smallskip%
\begin{tabular}{l}}%
{\end{tabular}%
\algcap%
\end{figure}%
}%
     
\title{Problemas com modelagem algébrica: três exemplos para semestres iniciais}

\author{João Batista S.\ de Oliveira\inst{1}}

\address{Faculdade de Informática -- Pontifícia Universidade Católica do Rio Grande do Sul
  (PUCRS)\\
  Av. Ipiranga, 6681 -- Porto Alegre -- RS
  \email{oliveira@inf.pucrs.br}
}

\begin{document} 

\maketitle

\begin{abstract}
  This paper describes three programming problems that are simple
  enough to be used in the beginning of a CS undergraduate program but
  also interesting enough to be worth exploring with different
  approaches. We are able to apply a mixture of programming
  practices, abstraction and algebraic approaches to the problems,
  so that these subjects may be presented as complementary and allowing
  for clear and elegant solutions.
\end{abstract}
     
\begin{resumo} 
  Este artigo descreve três problemas de programação que são simples o
  bastante para serem usados no início de um curso de computação e ao
  mesmo tempo interessantes o bastante para serem explorados de formas
  diferentes, como técnicas de programação, abstração e modelagem
  algébrica, de forma que estes assuntos possam ser apresentado como
  complementares na busca de soluções simples, claras e elegantes.
\end{resumo}

\section{Introdução e motivação}

Em uma descrição abrangente, a computação busca modelar e propor
soluções através do uso de computadores para problemas de outras áreas
do conhecimento ou da própria computação e uma das tarefas essenciais
de um curso na área de computação é mostrar técnicas que podem ser
usadas nesta busca de modelos e soluções.

Por outro lado, uma experiência frequente é notar que aspectos
complementares desta busca de soluções costumam ser apresentados em
disciplinas isoladas, em semestres distantes e que conversam entre si
menos do que poderiam conversar. Como consequencia, a vantagem costuma
ser da disciplina que ocupar o espaço inicial, geralmente uma
disciplina de programação enquanto disciplinas que esperam mais
formalismo e maturidade serão apresentadas mais tarde. Isto tem a
óbvia desvantagem de isolar os assuntos e, para os métodos que vierem
depois, a de serem considerados técnicas secundárias ou menos
relevantes por estudantes que já sabem resolver o problema sendo
abordado, mesmo que de formas mais trabalhosas. Além disso, a
separação de assuntos também estimula o uso de descrições diferentes
para os mesmos modelos, o que não facilita a integração.

Neste artigo apresentamos três problemas simples o bastante para serem
abordados nos semestres iniciais de um curso de computação e que
permitem combinar diferentes técnicas de modelagem, fornecendo um bom
espaço de atuação a todas e mostrando como elas podem ser usadas em
conjunto. Os três problemas são interessantes, não-triviais (para um
estudante de semestres iniciais) e permitem abordagens de modelagem
que fornecerão soluções limpas e elegantes.

\section{Primeiro problema} \label{sec:prob1}

O primeiro e mais simples problema é retirado do repositório de
problemas de concursos de programação da Universidade de
Valladolid~\cite{uva} e denominado ``The Errant Physicist''. Ele conta
a história de um cientista que deseja escrever um programa capaz de
receber dois polinômios de duas variáveis (sempre $x$ e $y$) em uma
notação pouco legível e efetuar sua multiplicação, apresentando o
polinômio resultante de forma mais legível como no exemplo abaixo:
\begin{verbatim}
  Entrada 1: -yx8+9x3-1+y
  Entrada 2: x5y+1+x3
  Saida
   13 2    11      8      6    5 2    5     3      3
- x  y  - x  y + 8x y + 9x  + x y  - x y + x y + 8x  + y - 1
\end{verbatim}
Este problema foi originalmente proposto para uso em uma maratona de
programação juntamente com vários outros problemas, por isso pode-se
esperar que uma solução seja proposta sem grande demora, o que costuma
se confirmar em sala de aula. Condições adicionais do problema são as
seguintes:
\begin{itemize} \itemsep 0em
 \item Os expoentes são sempre números positivos.
 \item Termos comuns devem ser simplificados. Por exemplo,
 $40x^{2}y^{3}-38x^{2}y^{3}$ será reduzido a $2x^{2}y^{3}$. Termos com
 coeficiente zero são omitidos.
 \item Coeficientes iguais a 1 são omitidos, exceto no caso de uma
 constante 1 como no exemplo.
 \item Expoentes iguais a 1 são omitidos e potências $x^0$ e $y^0$ são
 omitidas.
 \item Sinais de \texttt{+} e \texttt{-} tem sempre um espaço em
 branco à frente e outro atrás.
 \item A saída é feita em duas linhas, uma para os expoentes e outra
 para as variáveis e coeficientes.
 \item Os termos devem ser escritos em ordem decrescente dos expoentes
 de $x$ e, em caso de empate, em ordem decrescente de expoentes de
 $y$.
\end{itemize}
O problema pode ser dividido em três etapas: leitura das duas entradas
fornecidas, cálculo do polinômio resultante (com simplificação) e em
seguida a impressão no formato exigido. Embora a leitura das entradas
e a posterior escrita do resultado sejam etapas interessantes, nos
concentraremos em calcular o produto pois esta costuma ser considerada
pelos estudantes a etapa mais sofisticada da solução e obter uma
versão elegante para esta etapa causa impacto bastante positivo.

Em busca de uma representação simples, percebe-se que os termos dos
polinômios da entrada podem ser representados como triplas informando
o coeficiente e os expoentes de $x$ e $y$ como abaixo:
\[ -yx8 \equiv -1*x^8*y \equiv (-1, 8, 1) \]
e cada polinômio da entrada pode ser representado como uma lista ou 
conjunto de termos:
\begin{eqnarray*}
P_1 & = & \{ (-1,8,1), (9,3,0), (-1,0,0), (1,1,0) \} \\
P_2 & = & \{ (1,5,1), (1,0,0), (1,3,0) \}
\end{eqnarray*}
Se até este ponto a representação dos polinômios é bastante simples e
intuitiva, uma vez que eles estejam disponívels chega-se à parte mais
interessante: o cálculo do resultado, quando os termos de cada
polinômio são multiplicados um a um e armazenados:
\begin{algorithm}
{\caption{\label{alg-prob1} Multiplicação dos polinômios de entrada.}}
Mult($P_1, P_2$) :
\I  $Res = \{\}$;
\I  para $t_1 \in P_1$ :
\II   para $t_2 \in P_2$ :
\III    $Res = Res \cup \{ t_1 \otimes t_2 \} $;
\I  retorna $Res$;
\end{algorithm}
Evidentemente, a operação $\otimes$ é a peça essencial para terminar o
cálculo e pode ser reduzida a uma operação algébrica sobre as tuplas
apresentadas:
\begin{eqnarray*}
t_1 & = & (c_1, x_1, y_1) \\
t_2 & = & (c_2, x_2, y_2) \\
t_1 \otimes t_2 & = & (c_1 * c_2, x_1 + x_2, y_1 + y_2)
\end{eqnarray*}
Depois de tirar proveito deste modelo algébrico e obter o polinômio
desejado, ainda é necessário voltar à implementação para um processo
de limpeza do polinômio obtido, retirando-se termos com coeficiente
zero e simplificando termos de mesmos expoentes. Em seguida uma etapa
de ordenação pode ser aplicada para impressão do resultado. Mesmo este
processo simples pode ser usado para explorar novas possibilidades:
\begin{itemize} \itemsep 0pt
  \item Um processo de limpeza sem ordenar os termos do resultado
  é simples mas terá desempenho $O(n^2)$ onde $n$ é o número de termos;
  \item Ordenar os termos de acordo com os expoentes faz com que
  termos de mesmos expoentes estejam agrupados, o que facilita a
  simplificação mas exige que dois termos sejam retirados do resultado
  e um novo com a simplificação seja inserido na posição correta;
  \item Em qualquer caso, o processo de ordenação permite a
  construção de um operador $\leq$ sobre as tuplas propostas,
  explorando mais um aspecto da representação sugerida.
\end{itemize}
Mesmo uma outra abordagem para este problema fará essencialmente as
mesmas tarefas descritas acima pois ele não apresenta uma forma mais
simples de solução, mas é muito interessante deixar que os estudantes
resolvam o problema, fazendo suas descobertas e depois apresentar a
solução em forma algébrica acima dando uma nova clareza à solução
obtida e apresentando uma forma alternativa de descrição.

Além de ser simples, razoavelmente interessante e trazer a motivação
de estar resolvendo uma tarefa aparentemente complexa, o problema tem
outras duas vantagens: uma delas é a disponibilidade de teste dsa
implementações através do site original, pois isto oferece a
possibilidade de validação das soluções, bem como acostuma estudantes
a exigências de formato de entrada e saída muitas vezes ignoradas em
sala de aula ou consideradas pouco importantes.

Outro aspecto interessante é a modelagem com técnicas diferentes em
momentos diferentes. Listas e tuplas fornecem as estruturas básicas,
uma operação algébrica é introduzida para modelar o produto de termos
e outras operações entre listas são necessárias novamente para limpeza
e impressão, podendo também ser abordadas de maneira mais formal se
desejado. Uma possibilidade de exploração deste problema pode ser a
inclusão de mais variáveis ou a divisão entre polinômios, por exemplo.

\section{Segundo problema} \label{sec:prob2}

O segundo problema tem a virtude de ser aparentemente mais simples e
tem pelo menos duas soluções imediatas, uma delas com dois defeitos
graves e a outra com (apenas) um defeito grave. A terceira solução,
obtida com uma modelagem mais cuidadosa, oferece uma solução muito
eficiente. O enunciado é:
\begin{quote}
\it 
Seja a função $F_n$, que serve para produzir strings:
\[
  F_n=\left\{ \begin{array}{ll}
    A, & n=0\\
    B, & n=1\\
    F_{n-1}\oplus F_{n-2,} & n\geq2,
    \end{array}\right.
\]
onde o sinal $\oplus$ representa a concatenação de strings. Por
exemplo, 
\begin{eqnarray*}
  F_0 & = & A\\
  F_1 & = & B\\
  F_2 & = & BA\\
  F_3 & = & BAB\\
  F_4 & = & BABBA
\end{eqnarray*}
Deve-se determinar quantas vezes uma string $S$ com até 20 caracteres
está repetida dentro de uma string $F_n$, sabendo que $n\leq 50$.
Por exemplo, quantas vezes a string $AB$ está em
$F_{37}$?\footnote{14930352 vezes.}
\end{quote}

\subsection{Primeiras alternativas}

Para um estudante de semestres iniciais, a primeira solução é
inevitavelmente tentadora: construir a string $F_n$ desejada e depois
contar as ocorrências de $S$ dentro dela. É sempre uma experiência
interessante perceber como a certeza vitoriosa ao obter a resposta
para casos simples altera-se ao tentar tratar casos maiores onde ela
falha por falta de memória e/ou excessivo tempo de execução. Embora
decepcionante, essa primeira experiência traz a lição de que a solução
para um problema simples não precisa ser tão simples quanto o problema
e que soluções nem sempre são universais.

Depois de convencidos da impraticalidade de construir a string, uma
segunda solução se propõe: como a string $S$ a ser procurada tem no
máximo 20 caracteres é possível programar a função $F_n$
recursivamente e ao chegar a um de seus casos-base (sempre $A$ ou $B$)
concatena-se esta letra a um buffer $S'$ que inicia vazio e vai sendo
aumentado até ter o mesmo tamanho de $S$, retirando-se sua primeira
letra se preciso. A cada nova letra concatenada, se $S'=S$ temos uma
nova ocorrência de $S$ dentro de $F_n$. Esta abordagem é muito
interessante, certamente um passo significativo quando considerada
como técnica de resolução e embora não sofra mais com falta de memória
sofre do mesmo grave problema de desempenho e uma solução efetiva
exige uma modelagem mais cuidadosa.

\subsection{Outra possibilidade de modelagem}

Novas possibilidades surgem quando explora-se a construção de uma das
strings $F_n$. Por exemplo, para $F_{32}$ sabemos que
\[ F_{32} = F_{31} \oplus F_{30} \]
Embora não pareça muito rico, um exemplo concreto auxilia a constatar
alguns fatos:
\begin{itemize} \itemsep 0pt
  \item O número de ocorrências de $S$ em $F_{32}$ é tão grande quanto
  os números de ocorrências em $F_{31}$ e $F_{30}$ somados, já que
  ocorrências não podem ser retiradas.
  \item Se o número de ocorrências for maior do que esta soma é por que
  novas strings $S$ foram criadas na junção entre as duas strings.
  \item A junção onde podem surgir novas ocorrências de $S$ é composta
  pelos caracteres finais de $F_{31}$ e os caracteres iniciais de
  $F_{30}$.
  \item Na junção, para que tenhamos uma nova ocorrência de $S$ é
  preciso que um dos lados forneça no mínimo um caractere e o outro
  forneça no máximo $|S|-1$ caracteres.
  \item Adicionalmente, a string resultante $F_{32}$ tem o mesmo
  início de $F_{31}$ e o mesmo final de $F_{30}$. 
\end{itemize}

Com estes fatos, percebe-se que usar recursão talvez não seja é a
melhor alternativa para achar o número de ocorrências de $S$ e no caso
de um programa iterativo precisamos apenas somar as ocorrências
existentes nas strings anteriores e adicionar as eventuais ocorrências
da área de junção. A partir daí, podemos propor que a string produzida
por $F_n$ não seja mais construída explicitamente, mas representada
por um novo tipo de objeto
\[ ( P_n, \#_n, S_n ) \]
onde $P_n$ e $S_n$ são o prefixo e o sufixo de $F_n$ respectivamente
(indo no máximo até $|S|-1$, portanto não ultrapassando 19 caracteres)
e $\#_n$ é a quantidade de ocorrências de $S$ em $F_n$. A partir daí,
os termos iniciais da recorrência podem ser representados por 
\begin{eqnarray*}
  A & \equiv & ( ``A", S==``A", ``" ) \\
  B & \equiv & ( ``B", S==``B", ``" )
\end{eqnarray*}
onde o operador $==$ equivale ao operador de igualdade em linguagem C,
retornando 0 se as strings forem diferentes ou 1 se forem iguais. Com
esta representação podemos propor uma operação algébrica entre dois
destes objetos, produzindo o objeto correspondente à string seguinte.
Se 
\begin{eqnarray*}
  F_{n-2} & = & ( P_{n-2}, \#_{n-2}, S_{n-2} ) \\
  F_{n-1} & = & ( P_{n-1}, \#_{n-1}, S_{n-1} ) \\
  F_{n}   & \equiv & F_{n-1} \oplus F_{n-2}
\end{eqnarray*}
então a operação $\oplus$ pode ser definida como
\[  F_{n-1} \oplus F_{n-2} = ( P_{n-1}, \#_{n-1} + \#_{n-2} + \cap( S_{n-1} P_{n-2} ), S_{n-2} ) \] 
onde $ \cap( S_{n-1} P_{n-2} ) $ é uma função auxiliar que conta
quantas ocorrências de $S$ estão contidas na concatenação de $S_{n-1}$
e $P_{n-2}$.

\subsection{Uma versão prática}

Embora colocado em uma boa direção, este problema não está
completamente (nem corretamente) modelado quando chega-se à operação
$\oplus$ apresentada: para uma implementação ainda é necessário
preencher detalhes e tomar decisões. Por exemplo, os prefixos e
sufixos dos casos iniciais da modelagem tem 0 ou 1 caractere e a
operação $\oplus$ não alterará seu comprimento, apenas repassa-os ao
próximo objeto. Além disso, uma condição que está implícita no modelo
é que prefixos e sufixos não se sobreponham. Ou seja, a operação
descrita acima ainda não funcionará corretamente e por isso abrem-se
possibilidades:
\begin{itemize} \itemsep 0pt
  \item Adaptar o modelo para iniciar com os prefixos e sufixos dados
  e fazê-los aumentar até atingirem um tamanho igual a $|S|-1$. Esta
  decisão tem a desvantagem de deixar o modelo mais complexo embora
  possa ser favorecida se deseja-se desenvolver um formalismo completo;
  \item Por outro lado, já que a string $S$ está limitada a no máximo
  20 caracteres, uma boa abordagem prática cria as strings $F_n$
  explicitamente até o caso em que elas sejam maiores do que o dobro
  do tamanho de $S$ e depois retira-se destas strings seus prefixos,
  sufixos e contadores de ocorrências, criando as tuplas e a
  partir daí usando a operação $\oplus$ sugerida.

  Esta abordagem mista tem a vantagem de manter o modelo simples e
  mostrar como técnicas diferentes podem ser combinadas em uma mesma
  tarefa, mantendo a simplicidade em cada etapa e evitando as
  complicações do uso de apenas uma delas.
\end{itemize}

\section{Terceiro problema} \label{sec:prob3}

\begin{figure}[ht]
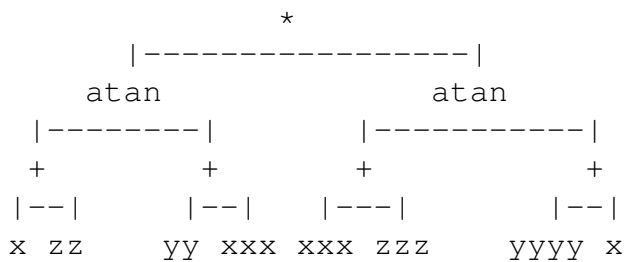

\lineskip -2mm
\begin{verbatim}
                   *
           |-----------------|
         atan              atan
      |--------|       |-----------|
      +        +       +           +
     |--|     |--|   |---|       |--|
     x zz    yy xxx xxx zzz    yyyy x
\end{verbatim}
\caption{Uma árvore representando uma expressão aritmética.}
\label{fig:figtree}
\end{figure}

O último e mais interessante problema a ser proposto diferencia-se dos
anteriores ao ter menos alternativas óbvias de solução e ao mesmo
tempo sua modelagem algébrica é bastante simples. A tarefa é imprimir
uma árvore binária correspondente a uma expressão aritmética em um
terminal, como no exemplo da Figura~\ref{fig:figtree}.
As condições do problema são as seguintes:
\begin{itemize} \itemsep 0pt
  \item Recebe-se inicialmente a referência para a raiz da árvore e
nenhuma informação adicional;
  \item Todos os nodos da árvore serão folhas ou terão dois filhos;
  \item Todo o nodo $n$ tem um label $label(n)$ e filhos $left(n)$ e
$right(n)$;
  \item O label de um nodo pode ter tamanho variável como no caso da
função de dois argumentos {\tt atan} no exemplo anterior;
  \item Assume-se que a versão impressa caberá na largura do terminal.
\end{itemize}

\begin{figure}[ht]
\centering
\includegraphics[width=.5\textwidth]{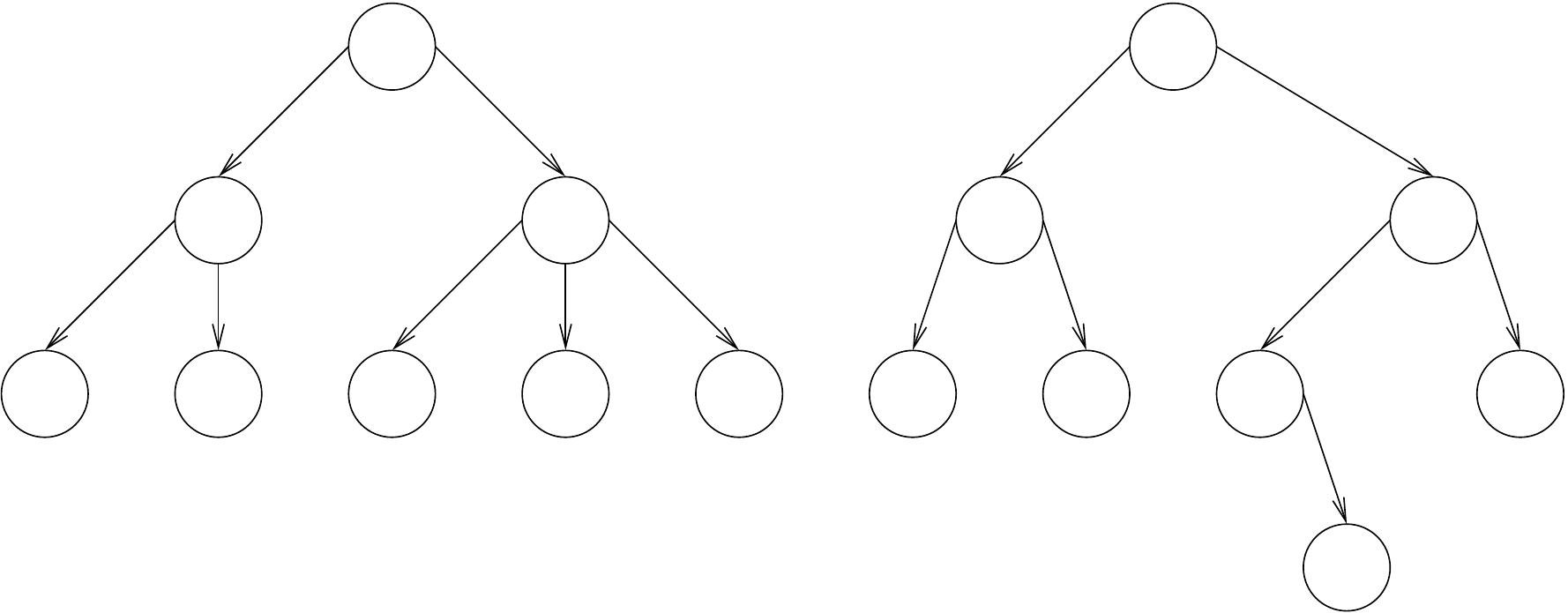}
\caption{Duas alternativas diferentes para o desenho de árvores.}
\label{fig:fig1}
\end{figure}

Existe uma variedade de técnicas diferentes para a tarefa, indo das
mais simples até variações bem mais complexas. Enquanto algumas
alternativas mais antigas usam algoritmos que podem produzir
resultados pouco satisfatórios ou que não tem muita
clareza~\cite{Wirth:1978:ADS:540029,SPE:SPE4380100706}, alternativas
posteriores~\cite{Walker:1990:NAG:79019.79026} já mostram um conjunto
mais bem estabelecido de critérios estéticos para um ``bom" desenho e
são capazes de imprimir árvores não-binárias. Mesmo assim, os
algoritmos apresentados ainda sofrem pelas notações e linguagens da
época. Em uma versão mais recente e menos ligada a uma linguagem, uma
boa descrição desta jornada algorítmica pode ser encontrada
em~\cite{mill_drawing_????}, que ao seu final apresenta um algoritmo
anterior~\cite{Buchheim02improvingwalker's} capaz de imprimir árvores
mas que sofre o efeito de já ser apresentado em uma linguagem
(Python). Em sua forma mais comumente aceita, espera-se que um
algoritmo para impressão de uma árvore seja capaz de imprimir uma
árvore genérica onde os labels de cada nodo possam ser sobrepostos
verticalmente, criando uma árvore que é o mais estreita possível como
na árvore à esquerda na Figura~\ref{fig:fig1}.

Para uma situação inicial, no entanto, optamos por restringir o
problema a árvores binárias sem permitir que nodos sejam sobrepostos
verticalmente, mesmo que sob estas condições ainda busquemos a
representação mais estreita possível e criando representações
similares à da árvore direita na Figura~\ref{fig:fig1}.
\subsection{Um modelo inicial}
Embora o desenho da Figura~\ref{fig:fig1} transmita a noção do que
desejamos obter, ele é enganoso em pelo menos um aspecto: os labels
colocados nos nodos não terão necessariamente o mesmo comprimento e
isto deve ser considerado no posicionamento. Além do label, para
reproduzir a Figura~\ref{fig:figtree} também é preciso posicionar a
barra horizontal que conecta um nodo aos filhos e que terá coordenadas
diferentes das coordenadas do label.

\begin{figure}[ht]
\centering
\includegraphics[width=.7\textwidth]{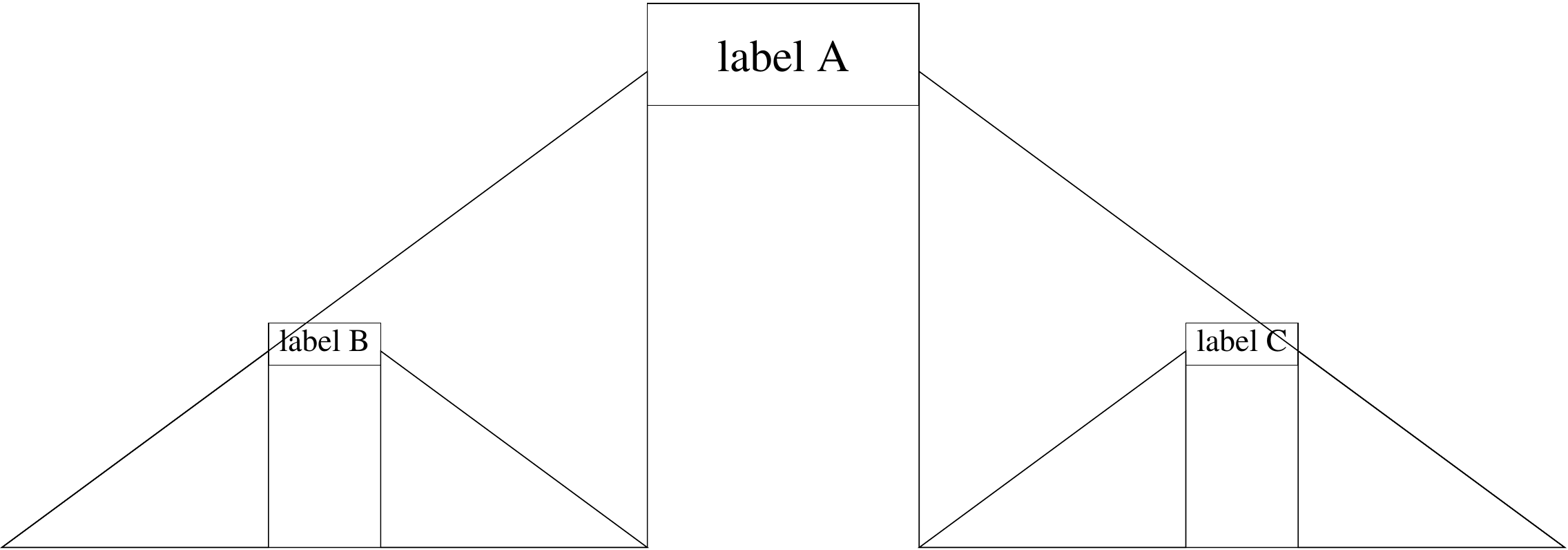}
\caption{Um modelo abstrato para o desenho.}
\label{fig:fig2}
\end{figure}

\begin{figure}[ht]
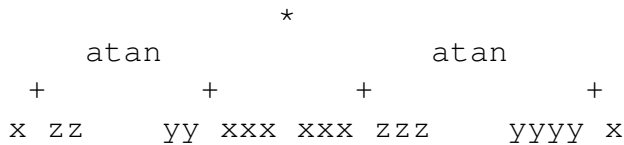

\begin{verbatim}
                   *
         atan              atan
      +        +       +           +
     x zz    yy xxx xxx zzz    yyyy x
\end{verbatim}
\caption{O primeiro resultado obtido, ainda sem as barras horizontais.}
\label{fig:figtree1}
\end{figure}

Como essencialmente é preciso decidir linha $l_n$ e coluna $c_n$ para
apresentação na tela de cada nodo $n$, é fácil concluir que a linha
adequada para cada nodo pode ser obtida com um caminhamento recursivo
que preenche $l_n$:
\begin{algorithm}
{\caption{\label{alg1-prob3} Preenchimento de $l_n$.}}
void Preenche($n$, $niv$) :
\I  se $n = NULL$ retorna;
\I  $l_n = niv$;
\I  Preenche( $left(n)$, $niv+1$ );
\I  Preenche( $right(n)$, $niv+1$ );
\end{algorithm}
O verdadeiro desafio está no cálculo das coordenadas horizontais para
cada nodo, e neste momento o tipo de desenho que desejamos pode ser
abstraído pela Figura~\ref{fig:fig2}.
Ela sugere que seja necessário conhecer a largura de toda a sub-árvore
a partir de um nodo $n$ para que o nodo pai de $n$ possa ser
posicionado corretamente, mas também sugere que seja útil ter este
valor separado em três valores:
\begin{itemize} \itemsep 0pt
  \item $esq(n)$ e $dir(n)$ guardam as larguras das sub-árvores esquerda e direita de $n$;
  \item $mid(n)$ guarda a largura da parte central de $n$;
\end{itemize}
É possível imaginar que o cálculo destes valores seja recursivo e deve
ser feito para os filhos e depois para o nodo pai, que terá seus
valores afetados pelos valores dos filhos, e isso encaixa-se
naturalmente no algoritmo esboçado: se precisamos calcular uma tripla
$T_n = (esq(n), mid(n), dir(n))$ para cada nodo $n$, temos o seguinte
modelo sugerido pela Figura~\ref{fig:fig2}:
\begin{eqnarray*}
  T_B & = & ( esq(B), mid(B), dir(B) ) \\
  T_C & = & ( esq(C), mid(C), dir(C) ) \\
  T_A & \equiv & T_B \wedge T_C
\end{eqnarray*}
onde a operação $\wedge$ carrega toda a responsabilidade por um
cálculo correto. Felizmente a própria Figura~\ref{fig:fig2} sugere
como deve ser feita a operação:
\[ T_B \wedge T_C \equiv ( \sum T_B, size( label(A) ), \sum T_C ) \]
E assim podemos obter uma versão mais completa do algoritmo inicial:
\begin{algorithm}
{\caption{\label{alg2-prob3} Um algoritmo alterado para $l_n$ e $(esq(n), mid(n), dir(n))$.}}
Tupla Preenche($n$, $niv$) :
\I  se $n = NULL$ retorna Tupla(0, 0, 0);
\I  $l_n = niv$;
\I  $T_B$ = Preenche( $left(n)$, $niv+1$ );
\I  $T_C$ = Preenche( $right(n)$, $niv+1$ );
\I  retorna Tupla($\sum T_B, size( label(n) ), \sum T_C$);
\end{algorithm}
Usando este método para preencher as informações para cada nodo,
podemos em seguida fazer seu posicionamento com um novo caminhamento
recursivo que posiciona primeiro os nodos à esquerda e depois move o
nodo pai para a direita de acordo com o espaço usado, obtendo-se a
Figura~\ref{fig:figtree1} ainda sem os traços horizontais.

Uma variação interessante pode surgir com a percepção de que o papel
de $esq(n), mid(n)$ e $dir(n)$ é guardar, embora de forma fracionada,
a largura total da sub-árvore ocupada pelo nodo $n$. Isso sugere
testes com o novo algoritmo da Figura~\ref{alg3-prob3}, que calcula a
largura $wid_n$ da sub-árvore de forma explícita (e fornecendo menos
informação do que as três partes isoladamente).
\begin{algorithm}
{\caption{\label{alg3-prob3} Um algoritmo reduzido para $l_n$ e $wid_n$.}}
int Preenche($n$, $niv$) :
\I  se $n = NULL$ retorna 0;
\I  $l_n = niv$;
\I  $T_B$ = Preenche( $left(n)$, $niv+1$ );
\I  $T_C$ = Preenche( $right(n)$, $niv+1$ );
\I  $wid_n = T_B + size( label(n) ) + T_C$;
\I  retorna $wid_n$;
\end{algorithm}

\begin{figure}[ht]
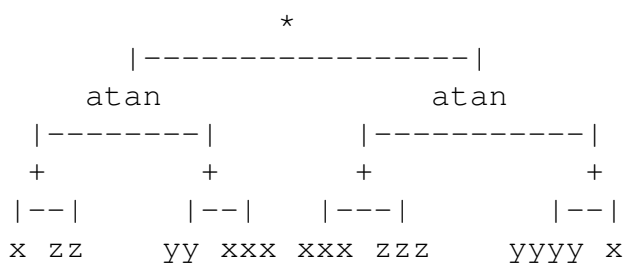

\begin{verbatim}
                   *
           |-----------------|
         atan              atan
      |--------|       |-----------|
      +        +       +           +
     |--|     |--|   |---|       |--|
     x zz    yy xxx xxx zzz    yyyy x
\end{verbatim}
\caption{Inclusão das barras horizontais. A árvore não é a mais estreita
possível, mas já é uma representação aceitável da original.}
\label{fig:figtree2}
\end{figure}

\begin{figure}[ht]
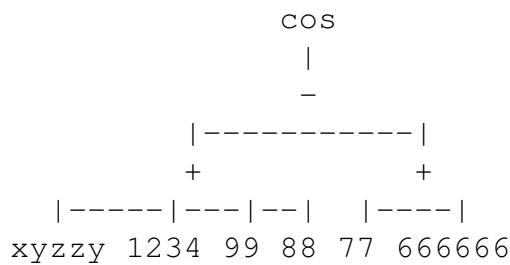

\begin{verbatim}
                        cos 
                         |
                         -
                   |-----------|
                   +           +
            |-----|---|--|  |----|
          xyzzy 1234 99 88 77 666666
\end{verbatim}
\caption{Árvore genérica e com largura mínima.}
\label{fig:figtree3}
\end{figure}

Para posicionar as barras abaixo do nodo $A$, podemos calcular onde
estão as coordenadas $b_e$ e $b_d$ das barras esquerda e direita
respectivamente usando as informações de seus filhos $B$ e $C$. Se o
nodo $A$ já deve ter um offset $of$ causado por nodos impressos
anteriormente, temos
\begin{eqnarray*}
  b_e & = & of + esq(B) + \lfloor mid(B)/2 \rfloor \\
  b_d & = & of + \sum T_B + size( label(A) ) + dir(C) + \lfloor mid(C)/2 \rfloor,
\end{eqnarray*}
o que produz a saída da Figura~\ref{fig:figtree2}.
Até este ponto o modelo apresentado é simples e bastante intuitivo,
com a vantagem de adaptar-se a nodos com labels de tamanhos diferentes
mas oferece pontos de exploração que não foram desenvolvidos neste
artigo:
\begin{itemize} \itemsep 0pt
  \item A árvore obtida na Figura~\ref{fig:figtree2} não é a mais
estreita possível. Por exemplo, as sub-árvores abaixo das operações
{\tt atan} tem entre elas um espaço correspondente ao tamanho do
rótulo do nodo pai, quando apenas um espaço em branco já seria
adequado. A busca pela menor largura possível altera a operação
$\wedge$ mudando não só o cálculo do componente central de $T_A$, mas
também dos componentes esquerdo e direito:
\begin{eqnarray*}
T_B \wedge T_C & \equiv ( & esq(B) + \lfloor mid(B)/2 \rfloor, \\
               &          & \max( \lceil mid(B)/2 \rceil + dir(B) + 1 + esq(C) + \lfloor mid(C)/2 \rfloor, size( label(A) ) ), \\
               &          & \lceil mid(C)/2 \rceil + dir(C) \quad )
\end{eqnarray*}
  \item Outra alteração possível é a generalização do modelo para uma
árvore não-binária. Realizando-se as mudanças necessárias na operação
$\wedge$ sugerida anteriormente, obteremos representações como a da
Figura~\ref{fig:figtree3}.
\end{itemize}
\section{Conclusão}

O objetivo deste artigo foi trazer exemplos de problemas envolvendo
alguma pergunta interessante a ser respondida, com facilidade de
compreensão do problema e algum tipo de solução que seja trabalhosa
mas possível já para um aluno dos semestres iniciais. Além disso eles
tem duas características importantes: grande simplificação conceitual
quando a modelagem da solução é feita com uma percepção algébrica do
problema e o fato de que uma solução completamente funcional pode
envolver esta modelagem elegante e outros aspectos que fazem parte do
dia-a-dia da programação: o uso eficiente de listas e outras
estruturas, decisões operacionais como ``ordenar e limpar'' ou
``limpar e ordenar'' e a exigência de entrada e saída em formatos
precisos, por exemplo.

\bibliographystyle{sbc}
\bibliography{a1}

\begin{thebibliography}{1}

\bibitem{uva}
Uva {Online} {Judge}, {https://uva.onlinejudge.org}.

\bibitem{Buchheim02improvingwalker's}
Christoph Buchheim, Michael J\"unger, and Sebastian Leipert.
\newblock Improving {W}alker's algorithm to run in linear time, 2002.

\bibitem{mill_drawing_????}
William Mill.
\newblock Drawing {Presentable} {Trees},
  {https://llimllib.github.io/pymag-trees}.

\bibitem{SPE:SPE4380100706}
Jean~G. Vaucher.
\newblock Pretty-printing of trees.
\newblock {\em Software: Practice and Experience}, 10(7):553--561, 1980.

\bibitem{Walker:1990:NAG:79019.79026}
J.~Que. Walker, II.
\newblock A node-positioning algorithm for general trees.
\newblock {\em Softw. Pract. Exper.}, 20(7):685--705, July 1990.

\bibitem{Wirth:1978:ADS:540029}
Niklaus Wirth.
\newblock {\em Algorithms + Data Structures = Programs}.
\newblock Prentice Hall PTR, Upper Saddle River, NJ, USA, 1978.

\end{thebibliography}

\end{document}